\documentclass[12pt]{iopart}
\usepackage{epsfig}
\usepackage{amsfonts}
\newcommand{\bee}{\begin{eqnarray}}
\newcommand{\eend}{\end{eqnarray}}

\begin{document}

\title{Electric field of a point-like charge in a strong magnetic field}
\author{A.E. Shabad$^*$ and V.V.Usov$^\dag$}$\\$
$^*$\textit{P.N. Lebedev Physics Institute, Moscow,
Russia}\eads{\mailto{shabad@lpi.ru}}$\\$ $^\dag$\textit{Center for
Astrophysics, Weizmann Institute of Science, Rehovot, Israel}
\eads{\mailto{fnusov@wicc.weizmann.ac.il}}
\begin{abstract} We describe the
potential produced by a point electric charge placed into a
constant magnetic field, so strong that the electron Larmour
length is much shorter than its Compton length. The standard
Coulomb law is  modified due to the vacuum polarization by the
external magnetic field. Only mode-2 photons mediate the static
interaction. The corresponding vacuum polarization component,
taken in the one-loop approximation, grows linearly with the
magnetic field. Thanks to this fact a scaling regime occurs in the
limit of infinite magnetic field, where the potential is
determined by a universal function, independent  the magnetic
field. 
The
scaling regime implies a short-range character of interaction in
the Larmour scale, expressed as a Yukawa law. On the contrary, the
electromagnetic interaction regains its long-range character in a
larger scale, characterized by the
 Compton length.  In this scale the tail of the Yukawa potential
 follows an anisotropic
Coulomb law: it  decreases  as the distance from the charge
increases, slower along the magnetic field and faster across. The
equipotential surface is an ellipsoid stretched along the magnetic
field. As a whole, the modified Coulomb potential is a narrower
function than the standard Coulomb function, the narrower the
stronger the field. The singular behavior in the vicinity of the
charge remains unsuppressed by the magnetic field. These results
may be useful for studying atomic spectra in super-strong magnetic
fields of several Schwinger's characteristic values.

\end{abstract}

\vspace{1cm}

\section{Introduction}

The fact that the vacuum, in which an external magnetic field is
present, is an optically  anisotropic medium, has been known,
perhaps, since the time when the nonlinearity of quantum
electrodynamics was first recognized: in a nonlinear theory
electromagnetic fields do interact with one another, provided the
strength of at least one of them is sufficiently strong (of the
order of or larger than $m^2/e$, where $m$ and $e$ are electron
mass and charge, respectively.) If the external field is strong,
other fields interact with it, the result of the interaction
depending upon the direction specified by the external field,
hence the anisotropy. Depending on the wave amplitude, the
electromagnetic waves propagation in this medium may be considered
as a nonlinear process \cite{bialyniska}, including the
transformation of one photon into two \cite{adler} or more
photons, or taken in the linear approximation with respect to the
amplitude. In the latter case, the second-rank polarization tensor
is responsible for the properties of the medium. In the kinematic
domain where the photon absorption processes like
electron-positron pair creation are not allowed, the polarization
tensor is symmetric and real, and the medium is transparent and
birefringent\footnote{If the scheme is extended to
electron-positron plasma in a magnetic field \cite{perez}, the
polarization tensor is no longer symmetric, but remains Hermitian
in the transparency domain. The absorption mechanism in this case,
apart from the pair creation, includes also the inverse Cherenkov
and inverse cyclotron radiation of the plasma electrons.}. In the
absorption domain the medium is dichroic \cite{heyl}. The limit of
low frequency and momentum belongs to the transparency domain and
corresponds to a constant anisotropic dielectric permeability of
the medium. In this limit the polarization tensor may be obtained
by differentiations with respect to the fields of an effective
Lagrangian, calculated on the class of constant external electric
and magnetic fields. For small values of these fields \cite{blp}
and for
 extremely large  \cite{heyl} fields the polarization operator
was in this way considered using the Heisenberg-Euler effective
Lagrangian calculated \cite{euler} within the one-loop
approximation. The knowledge of this limit is useful for studying
the dielectric screening of the fields that are (almost) static
and (almost) constant in space. For more general purposes,
however, this limit is not sufficient, and one should calculate
the polarization tensor directly, using the Feynman diagram
technique of the Furry picture in the external magnetic field. On
the photon mass shell, i.e. when the photon energy, $k_0$, and
3-momentum, $\bf k$, are related by the free vacuum dispersion law
$k_0^2={\bf k}^2,$ such calculations were done by S.Adler
\cite{adler2} and D.H.Constantinescu \cite{constantinescu}. The
results obtained are appropriate for handling the photon
propagation in weakly dispersive medium, when the dispersion law
does not essentially deviate from its vacuum shape. The
polarization operator for the case of general relation between the
photon mass and momentum was calculated by I.A.Batalin and
A.E.Shabad \cite{batalin}, W.Tsai \cite{tsai}, V.N.Baier et {\em
al.} \cite{baier}, D.V.Melrose and R.J.Stoneham \cite{melrose1}.
This gave the possibility of studying the photon propagation
\cite{annphys} under the conditions where the deviation from the
vacuum dispersion law may be very strong either due to the
phenomenon of the cyclotron resonance in the vacuum polarization
\cite{nuovcimlet} - this phenomenon is responsible for the effect
of photon capture by a magnetic field \cite{nature}, \cite{ShUs},
\cite{wunner} - or due to magnetic fields, much larger than the
characteristic value $B_0=m^2/e\simeq 4.4\times 10^{13}$G
\cite{kratkie}, \cite{shabtrudy}, \cite{mikheev}, \cite{japan}, or
due to the both circumstances \cite{zhetf}.

Now we start an investigation of problems of electro- and
magneto-statics in the presence of a strong external magnetic
field in the vacuum. For this purpose  expressions for the
dielectric permeability of the magnetized vacuum obtained from the
Heisenberg-Euler Lagrangian are applicable only as far as  the
fields slowly varying in the space are concerned. Otherwise, the
spatial dispersion becomes important. For this reason, when
considering the electric field produced by a point-like electric
charge in the present paper, we refer again to the polarization
tensor calculated off-shell in (\cite{batalin})-(\cite{annphys}),
taking it in the static limit $k_0=0$, but keeping the dependence
on $\bf k$.

Using the tensor decomposition of the polarization operator and
the photon Green function, established in \cite{batalin},
\cite{annphys} in an approximation-independent way, we find that
photons of only one polarization mode (mode-2 in the nomenclature
of these references) may be carriers of electrostatic force. This
is in agreement with the fact that the electromagnetic field of
these photons is, in the static limit, purely electric and
longitudinal. The photons of the other two modes mediate in this
limit the magneto-static field of constant currents.

To describe the static field, produced in the magnetized vacuum by
a point electric charge, we confine ourselves to considering the
asymptotically large magnetic fields $B\gg B_0$, such that the
electron Larmour length $(eB)^{-1/2}$ is much less than the
electron Compton length $m^{-1}$. Therefore, two different scales
occur in the problem: the Larmour scale and the Compton scale. A
simplifying expression for the mode-2
 eigenvalue of the polarization operator is used, valid for such fields.
 It was first
obtained by Yu.M.Loskutov and V.V.Skobelev \cite{skobelev} within
a special two-dimensional  technique intended for large fields,
and by A.E.Shabad \cite{kratkie}, and D.B.Melrose and R.J.Stoneham
\cite{melrose1} as the asymptote of the mode-2 eigenvalue
calculated (\cite{batalin})-(\cite{annphys}) in the one-loop
approximation. (See \cite{zhetf} for the detailed derivation of
the large-field asymptotic behavior.) The most important, now well
known, fact about this asymptotic behavior (see, {\em e.g.,} the
monographs \cite{dittrich}, \cite{shabtrudy},  \cite{kuznetsov})
is that the mode-2 eigenvalue contains a term linearly growing
with the magnetic field. It is sometimes expected that this term -
it appears in the denominator of the photon propagator and hence
of the expression for the potential - should lead to suppression
of the interaction mediated by mode-2 photons. In a different
problem we already had an opportunity to show that this is not
exactly the case \cite{prl}, \cite{prd}. We demonstrate in the
present paper that
 this term is crucial for the possibility that a scaling regime
might be achieved in the limit of an infinite magnetic field,
characterized by a comparatively simple universal function,
independent of the magnetic field. This function gives the
potential of a point charge in the energy units of $(eB)^{1/2}$ as
a dimensionless function of the space coordinates in the units of
the electron Larmour length $(eB)^{-1/2}$. Except for the closest
vicinity of the source charge, its form coincides with the Yukawa
law (see Eq.(\ref{yuk}) below) with the dimensionless mass
parameter $2\alpha/\pi$ (which is the "effective photon mass"
discussed in \cite{kuznets}, measured in the
 units $(eB)^{1/2}$). Thus, this mass governs the  exponential
 decrease of the potential away from its source. As one moves
 farther from the source, the Yukawa decrease ceases, and the
 potential tail approaches the anisotropic Coulomb shape of the
 form of Eq.(\ref{llargex}). The law of
 decreasing along the magnetic field is unaffected by the latter,
 the decrease is the fastest in the direction orthogonal to the
 magnetic field. Therefore, the linearly growing term in the mode-2
 eigenvalue of the polarization operator leads, first, to the faster
 decrease of the potential in the direction across the magnetic
 field for large distances, and, second, to its steeper shape for small
 distances. This may be recognized as suppression, indeed. On the other
 hand, the long-distance behavior along the magnetic field, as
 well as the standard Coulomb singularity near the source do not
 sense the magnetic field at all, no matter how strong it is.

In Section 2 we give an approximation-independent derivation
 for the form of the potential of a point charge in terms of the
mode-2 component of the photon propagator. In Section 3 we
consider the short-distance behavior of the potential. Three terms
of its asymptotic expansion near the source are explicitly
written. The universal function for the scaling infinite-field
limit is obtained as a one-fold integral. In Section 4 the
anisotropic Coulomb law is obtained for large distances by
considering different mathematical means, appropriate for
different remote regions in the space. The analytical results for
the potential are supplemented by its shapes drawn by a computer,
presented in Figures 1 - 5. 

\section{General representation for the static potential of a point-like  charge}
Electromagnetic 4-vector potential produced by the 4-current
$j_\nu(y)$ is \bee\label{4-pot} A_\mu(x)=\int
D_{\mu\nu}(x-y)j^\nu(y)\rmd^4y,\quad \mu,\nu=0,1,2,3.\eend Here
$x$ and $y$ are 4-coordinates, and $D_{\mu\nu}(x-y)$ is the photon
Green function in a magnetic field in the coordinate
representation. The metric in the Minkowski space is defined so
that diag $g_{\mu \nu}=(1,-1,-1,-1)$. Eq. (\ref{4-pot}) defines
the 4-vector potential with the arbitrariness of a free-field
solution, including the gauge arbitrariness. If the photon Green
function is chosen as causal, only the gauge arbitrariness
remains.

The current of a point-like static charge $q$, placed in the point
${\bf y}=0$ is\bee\label{current} j^\nu(y)=q\delta_{\nu
0}\delta^3({\bf y}).\eend Hence\bee\label{4-pot2} A_\mu({\bf
x})=q\int_{-\infty}^\infty D_{\mu 0}(x_0-y_0,{\bf x})\rmd y_0\nonumber\\
=q\int_{-\infty}^\infty  D_{\mu 0}(x_0+y_0,{\bf x})\rmd
y_0=q\int_{-\infty}^\infty D_{\mu 0}(y_0,{\bf x})\rmd y_0.\eend
This 4-vector potential is also static.

Where there is no magnetic field, and the photon propagator is
free and taken in the Feynman gauge (with the pole handled in a
causal way) \bee\label{free}D_{\mu\nu}(x-y)=D_{\mu\nu}^{\rm C}
(x-y)\equiv\frac {g_{\mu\nu}}{\rmi 4\pi^2(x-y)^2},\eend  only the
zeroth component of the 4-vector potential is
present:\bee\label{free2} A_0^{\rm C}({\bf
x})=q\int_{-\infty}^\infty D^{\rm C}_{\mu 0}(y_0,{\bf x})\rmd
y_0=\frac{q}{\rmi 4\pi^2}\int_{-\infty}^\infty \frac{\rmd
y_0}{y_0^2-{\bf x}^2}=\frac 1{4\pi}\frac{q}{|{\bf x}|}.\eend This
is the Coulomb potential in the Heaviside-Lorentz system of units
used throughout.

Let there be an external magnetic field $B$ directed along axis 3
in the Lorentz frame where the charge $q$ is at rest in the origin
${\bf x}=0$ , and no external electric field exists in this frame.
Call this frame special. Define the Fourier transform as
\begin{eqnarray}\label{fourier} D_{\mu\nu}(x)=\frac
1{(2\pi)^4}\int \exp ({\rm i}kx) D_{\mu\nu}(k)~{\rm d}^4k,\quad
\mu,\nu=0,1,2,3.
\end{eqnarray} Then (\ref{4-pot2}) becomes\begin{eqnarray}\label{potmag}
\hspace{-3cm} A_\mu({\bf x})=\frac{q}{(2\pi)^4}\int \exp ({\rm
i}(k_0y_0-{\bf kx})) D_{\mu 0}(k)~{\rm d}^4k\rmd y_0 =
\frac{q}{(2\pi)^3}\int D_{\mu 0}(0,{\bf k})\exp(-\rmi{\bf
kx})\rmd^3k.
\end{eqnarray}

The four 4-eigenvectors $b^{(a)}_\nu$, $a=1,2,3,4$, of the
polarization tensor $\Pi_{\mu\nu}$ \cite{batalin, annphys,
nuovcimlet} take in the special frame (and arbitrary
normalization) the form - the components are counted downwards as
$\nu= 0,1,2,3$ -
\begin{eqnarray}\label{b}
b_\nu^{(1)}=k^2\left(\begin{tabular}{c}0\\$k_1$\\$k_2$\\0\end{tabular}\right)_\nu+
k_\perp^2\left(\begin{tabular}{c}$k_0$\\$k_1$\\$k_2$\\$k_3$\end{tabular}\right)_\nu,\quad
b_\nu^{(2)}=\left(\begin{tabular}{c}$k_3$\\0\\0\\$k_0$\end{tabular}\right)_\nu,\nonumber\\
b_\nu^{(3)}=\left(\begin{tabular}{c}0\\$k_2$\\$-k_1$\\0\end{tabular}\right)_\nu,\quad
b_\nu^{(4)}=\left(\begin{tabular}{c}$k_0$\\$k_1$\\$k_2$\\$k_3$\end{tabular}\right)_\nu,
\end{eqnarray} Among them there is only one, whose zeroth component 
survives the substitution $k_0=0$. It is $b_\nu^{(2)}$. This
implies that out of the four ingredients of the general
decomposition of the photon
propagator\begin{eqnarray}\label{decomposition} D_{\mu\nu}(k)=
\sum_{a=1}^4 D_a(k)~\frac{b_\mu^{(a)}~
b_\nu^{(a)}}{(b^{(a)})^2},\nonumber\\
D_a(k)=\left\{\begin{tabular}{cc}$-(k^2+\kappa_a(k))^{-1},$
&$\quad a$=1,2,3\\arbitrary, &$a$=4\end{tabular}\;,\right.
\end{eqnarray}where $\kappa_a(k)$ are scalar eigenvalues of the
polarization tensor:
\begin{eqnarray}\label{eigen}\Pi_\mu^{~\nu}(k)~b^{(a)}_\nu=\kappa_a(k)~b^{(a)}_\mu,
\qquad \kappa_4(k)=0,\end{eqnarray} only the term with $a=2$,
$D_2(k)b_{\mu}^{(2)}b_{\nu}^{(2)}/(b^{(2)})^2$, participates in
(\ref{potmag}), i.e. only  mode-2 (virtual) photon may be a
carrier of electro-static interaction, and not photons of modes
1,2, nor the purely gauge mode 4. Bearing in mind that
$(b^{(2)})^2=k_3^2-k_0^2$, we have\bee\label{A0} A_0({\bf
x})=\frac{q}{(2\pi)^3}\int\frac{\rme^{-\rmi{\bf kx}}\rmd^3k}{{\bf
k}^2-\kappa_2(0,k_3^2,k_\perp^2)},~~A_{1,2,3}({\bf x})=0.\eend
Here $k_\perp^2=k_1^2+k_2^2$. Thus, the static charge gives rise
to electric field only, as it might be expected. The gauge
arbitrariness in the choice of the photon propagator
$D_4(k)b_{\mu}^{(4)}b_{\nu}^{(4)}=D_4(k)k_{\mu}k_{\nu}$ indicated
in (\ref{decomposition}) has no effect in (\ref{A0}). Certainly,
the potential (\ref{A0}) is defined up to gauge transformations.

The result that only mode-2 photons mediate electrostatic
interaction may be understood, if we inspect electric and magnetic
components of the fields of the eigen-modes obtained from their
4-vector potentials (\ref{b}) in the standard way: ${\bf
e}^{(a)}=k_0{\bf b}^{(a)}-{\bf k}b_0^{(a)}$, ${\bf h}^{(a)}={\bf
k}\times{\bf b}^{(a)}.$ These are \cite{annphys} \bee\label{5}
{\bf e}^{(1)} =-\frac{{\bf k_\perp}}{k_\perp}k_0, \hspace{35mm}
{\bf h}^{(1)} = (\frac{\bf k_\perp}{k_\perp} \times {\bf k}_3),
\eend\bee\label{5b} {\bf e}^{(2)}_\perp = {\bf k}_\perp k_3,
\hspace{7mm}{\bf e}^{(2)}_3 = \frac{{\bf k}_3}{k_3}
(k_3^2-k_0^2),\hspace{3mm} {\bf h}^{(2)} = -k_0\left({\bf
k_\perp}\times \frac{{\bf k_3}}{k_3}\right),\eend\bee\label{5c}
{\bf e}^{(3)} = - k_0 \left(\frac{{\bf
k_\perp}}{k_\perp}\times\frac{{\bf
k_3}}{k_3}\right),\hspace{3mm}{\bf h_\perp}^{(3)}=-\frac{\bf
k_\perp}{k_\perp} k_3,\hspace{3mm} {\bf h_3}^{(3)}= \frac{\bf
k_3}{k}_3 k_\perp.  \eend Here the cross stands for the vector
product, and the boldfaced letters with subscripts $3$ and $\perp$
denote vectors along the directions, parallel and perpendicular to
the external magnetic field, resp.

The photon energy and momenta here are not, generally, connected
by any dispersion law. Therefore, we can discuss polarizations of
virtual photons - carriers of the interaction - basing on Eqs.
(\ref{5})-(\ref{5c}). The electric field $\bf e$ in mode 1 is
parallel to $\bf k_\perp$, in mode 2 it lies in the plane
containing the vectors $\bf k, B$, in mode 3 it is orthogonal to
this plane, $i.e.$ mode 3 is always transversely polarized, ${\bf
e}^{(3)}{\bf k}=0$.   For the special case of the virtual photon
propagation transverse to the external magnetic field, $k_3=0$,
(this reduces to the general case of propagation under any angle
$\theta\neq 0$ by a Lorentz boost along the external magnetic
field), mode 2 is transversely polarized , ${\bf e}^{(2)}{\bf
k}=0$,  as is always the case for mode 3. Mode 1 for transverse
propagation, $k_3=0$, is longitudinally polarized, ${\bf}
e^{(3)}\times {\bf k}=0$, and its magnetic field is zero. The
lowest-lying cyclotron resonance of the vacuum polarization
\cite{nuovcimlet}, the one that corresponds to the threshold
$k_0^2-k_3^2=4m^2$ of creation of the pair of electron and
positron in the lowest Landau state each, belongs to mode 2. It
gives rise to the photon capture effect with the photon turning
into a free \cite{nature} or bound \cite{ShUs, wunner}
electron-positron pair. Another consequence of the cyclotron
resonance is that a real photon of mode 2 undergoes the strongest
refraction in the large magnetic field limit \cite{zhetf} even if
its frequency is far beyond the pair production threshold.

In the static limit $k_0=0$ the magnetic field in mode 2
disappears, ${\bf h}^{(2)}=0$, while its electric field is
collinear with ${\bf k},$ ${\bf e}^{(2)}={\bf k}.$ It becomes a
purely longitudinal virtual photon. Unlike  mode 2, in modes 1 and
3 in the static limit $k_0=0$ only the magnetic fields survive:
${\bf e}^{(1,2)}=0$, ${\bf h}^{(1)}={\bf k}_\perp\times{\bf B},$
${\bf h}_\perp^{(3)}=-{\bf k}_\perp k_3,$ $h_3^{(3)}=k_\perp^2$,
${\bf h}^{(1,3)}{\bf k}=0$. (Here normalizations are arbitrary and
kept fixed only within the same mode). A virtual mode-1 photon
carries magneto-static interaction. It is responsible for magnetic
field produced by a current flowing through a straight-linear
conductor oriented along the external magnetic field. In
accordance with the above formula for ${\bf h}^{(1)}$ its magnetic
field is orthogonal both to ${\bf B}$ and to the radial direction
in the transverse plane ${\bf k}_\perp$, along which the magnetic
field of the current decreases. The mode-3 photon contributes as
an interaction carrier in the problem of a magneto-static field
produced by a solenoid with its axis along axis 3. In the present
paper, however, we do not consider magneto-static problems.

In the asymptotic limit of high magnetic field $B\gg k_3^2,~~B\gg
m^2/e\equiv B_0$ the eigenvalue $\kappa_2(0,{\bf k}),$ as
calculated within the one-loop approximation \cite{batalin},
\cite{annphys}, with the accuracy of terms that grow with $B$ only
as its logarithm and slower is \cite{kratkie}
\begin{eqnarray} \kappa_2(0,k_3^2,k_\perp^2) =-\frac{2\alpha
bm^2}{\pi }\exp \left(- \frac{k_\bot^2}{2m^2b}
\right)T\left(\frac{k_3^2}{4m^2}\right),\label{2}\end{eqnarray}
where $b=B/B_0$ and\begin{eqnarray}
T(y)=y\int_{0}^1\frac{(1-\eta^2)\rm d \eta}{1+y(1-\eta^2)}=1-\frac
1{2\sqrt{y(1+y)}}\ln\frac{\sqrt{1+y}+\sqrt{y}}{\sqrt{1+y}-\sqrt{y}}
.\label{T}\end{eqnarray} Note that $T(y\rightarrow 0)\sim
2y/3,\;T(\infty)=1$.

Other eigenvalues, $\kappa_{1,3},$ do not contain the coefficient
$b$ that provides the linear increase of  $\kappa_2$ (\ref{2})
with the magnetic field.

 Expression for $\kappa_2$ (\ref{2}) is to
be used in (\ref{A0}) or, equivalently, in the expression
\bee\label{bessel}A_0({\bf x})=\frac{q}{2(2\pi)^2}\int_0^\infty
J_0(k_\perp x_\perp)\left(\int_{-\infty}^\infty\frac{\rme^{-\rmi
k_3x_3}\rmd k_3}{
k_\perp^2+k_3^2-\kappa_2(0,k_3^2,k_\perp^2)}\right)\rmd
k_\perp^2\eend obtained from (\ref {A0}) by integration over the
angle between the 2-vectors ${\bf k}_\perp$ and ${\bf x}_\perp$,
which are projections of ${\bf k}$ and ${\bf x}$ onto the plane
transverse to the magnetic field. In (\ref{bessel})
$x_\perp=\sqrt{x_1^2+x_2^2}$ and $J_0$ is the Bessel function of
order zero.

The nontrivial - other than the leading asymptote $\sim k_3^2$
near the point $k_3=k_\perp =0$ - dependence of the polarization
operator eigenvalue (\ref{2}) on the photon momentum components
$k_3$, $k_\perp$ is the spatial dispersion. We shall see in
Section 3 that it is important in the vicinity of the charge,
where the field has a large gradient. As for the anisotropic
behavior far from the charge, to be studied in Section 4, only the
above asymptote is essential, inferable also from the
Heisenberg-Euler Lagrangian.

\section{Short-distance behavior}
\subsection{Asymptotic expansion}
 To consider the
behavior of the potential  near its point-like source let us add
to and subtract from (\ref{bessel}) the standard Coulomb potential
(\ref{free2}) in the form \bee\label{shortperp}A_0^{\rm C}({\bf
x})=\frac q{(2\pi)^3}\int\frac{\rme^{-\rmi{\bf kx}}\rmd^3k}{{\bf
k}^2}=\nonumber\\=\frac {q}{2(2\pi)^2}\int_0^\infty J_0(k_\perp
x_\perp)\left(\int_{-\infty}^\infty\frac{\rme^{-\rmi k_3x_3}\rmd
k_3}{ k_\perp^2+k_3^2}\right)\rmd k_\perp^2=\frac 1{4\pi}\frac
q{\sqrt{x_\perp^2+x_3^2}}\eend so that \bee\label{difference1}
A_0({\bf x})=A_0^{\rm C}({\bf x})-\Delta A_0({\bf x}), \eend where
\bee\label{difference2} \hspace{-1cm}\Delta A_0({\bf
x})=\nonumber\\
\hspace{-1cm}\frac{q}{2(2\pi)^2}\int_0^\infty J_0(k_\perp
x_\perp)\int_{-\infty}^\infty\left(\frac{\rme^{-\rmi k_3x_3}}{
k_\perp^2+k_3^2}-\frac{\rme^{-\rmi k_3x_3}}{
k_\perp^2+k_3^2-\kappa_2(0,k_3^2,k_\perp^2)}\right)\rmd k_3\rmd
k_\perp^2.\eend  Note that the function $\Delta A_0(x_3,x_\perp)$
is an entire function of $x_\perp$, since the exponential in
(\ref{2}) provides convergence of the integral (\ref{difference2})
for any complex value of this variable. 
Keeping quadratic terms in the power series expansion of
$J_0(k_\perp x_\perp)$ and $\exp (-\rmi k_3x_3)$ in
(\ref{difference2}) we obtain  the first three terms  of the
asymptotic expansion of the potential (\ref{bessel}) near the
origin $x_3=x_\perp=0$ \bee\label{expand}A_0({\bf x})\sim
\frac{q}{4\pi}\left(\frac 1 {|\bf
x|}-2m(C-(2mx_\perp)^2C_\perp-(2mx_3)^2C_\parallel)\right), \eend
where $C$, $C_\perp$ and $C_\parallel$ are  dimensionless positive
constants depending on the external
field:\bee\label{constant}\hspace{-1.5cm}C\equiv
\frac{2\pi}{qm}\Delta A_0(0)=\nonumber\\\hspace{-1cm}=\frac{\alpha
b m}{\pi^2}\int_0^\infty
T\left(\frac{k_3^2}{4m^2}\right)\int_0^\infty \frac
{\exp\left(-\frac{k_\perp^2}{2m^2b}\right) \rmd k_\perp^2}
{(k_\perp^2+k_3^2)(k_\perp^2+k_3^2-\kappa_2(0,k_3^2,k_\perp^2))}
\rmd k_3 ,\eend \bee\label{constantperp}\hspace{-1.5cm}C_\perp
=\frac{\alpha b }{16m\pi^2}\int_0^\infty
T\left(\frac{k_3^2}{4m^2}\right)\int_0^\infty \frac
{k_\perp^2\exp\left(-\frac{k_\perp^2}{2m^2b}\right) \rmd
k_\perp^2}
{(k_\perp^2+k_3^2)(k_\perp^2+k_3^2-\kappa_2(0,k_3^2,k_\perp^2))}
\rmd k_3 ,\eend  \bee\label{constantpar}\hspace{-1.5cm}C_\parallel
=\frac{\alpha b }{8m\pi^2}\int_0^\infty
T\left(\frac{k_3^2}{4m^2}\right)\int_0^\infty \frac
{k_3^2\exp\left(-\frac{k_\perp^2}{2m^2b}\right) \rmd k_\perp^2}
{(k_\perp^2+k_3^2)(k_\perp^2+k_3^2-\kappa_2(0,k_3^2,k_\perp^2))}
\rmd k_3 ,\eend Thanks to the exponential factor the integrals
over $k_\perp^2$ here are fast converging. The resulting functions
decrease for large $k_3$ as $\sim 1/k^4_3$, so the remaining
integrals over $k_3$ in (\ref{constant}), (\ref{constantperp}),
(\ref{constantpar}) converge, bearing in mind that $T$ is a
bounded function.
\begin{figure}[htb]
  \begin{center}
   \includegraphics[bb = 0 0 405 210,
    scale=1]{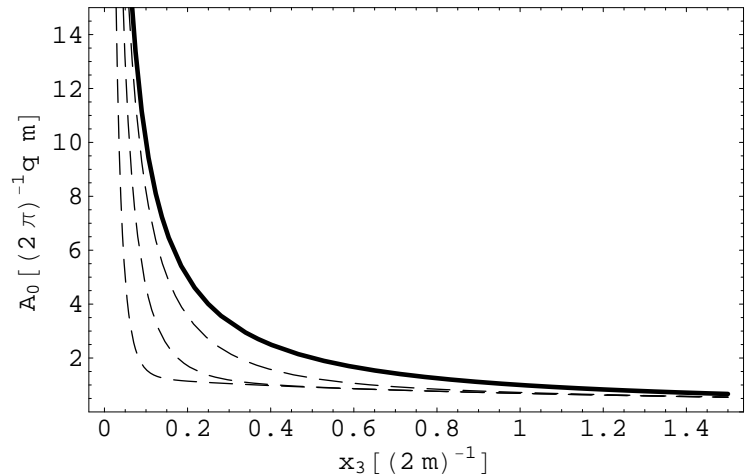}
\caption{Modified Coulomb potential $A_0(x_3,0)$ (\ref{bessel})
plotted along the axis $x_3$ passing through the charge $q$
parallel to the magnetic field. Dashed lines correspond to three
values of the magnetic field (from left to right): $B=10^6B_0,
B=10^5B_0$ and $B=10^4B_0$.  Thick solid line is the standard
Coulomb law (\ref{free2}) $A_0^{\rm C}(x_3,0)=q/4\pi x_3$. Dashed
lines approach asymptotically the solid line at the both edges of
the figure. The abscissa represents the distance along the
magnetic field in the units $(2m)^{-1}$. The ordinate represents
the potential in the units $qm/2\pi$.} \label{fig:1}
  \end{center}
\end{figure}
\begin{figure}[htb]
  \begin{center}
   \includegraphics[bb = 0 0 405 210,
    scale=1]{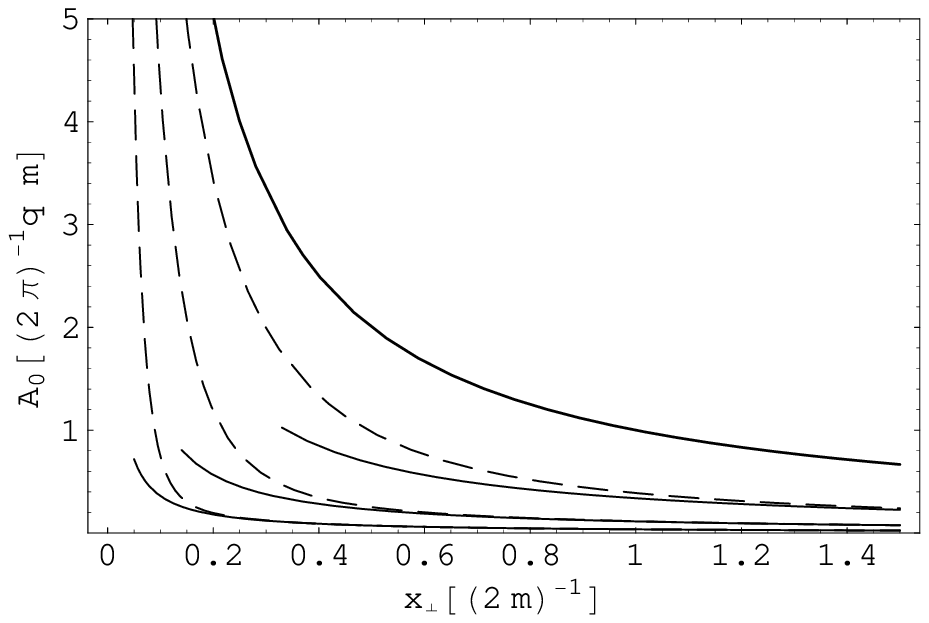}
\caption{Modified Coulomb potential $A_0(0,x_\perp)$
(\ref{bessel}) plotted along the axis $x_\perp$ passing through
the charge $q$ transversely to the magnetic field. Dashed lines
correspond to three values of the magnetic field (from left to
right): $B=10^6B_0, B=10^5B_0$ and $B=10^4B_0$. Solid line is the
standard Coulomb law (\ref{free2}) $A_0^{\rm C}(0,x_\perp)=q/4\pi
x_3$. Dashed lines approach asymptotically the solid line at the
left edge of the figure and the short solid lines (\ref{llargex})
$A_0(0,x_\perp)=q/4\pi x_\perp^\prime$ at the right edge. The
abscissa represents the distance across the magnetic field in the
units $(2m)^{-1}$. The ordinate represents the potential in the
units $qm/2\pi.$} \label{fig:2}
  \end{center}
\end{figure}
The values of the coefficients 
(\ref{constant}), (\ref{constantperp}), (\ref{constantpar})
calculated for four values of the magnetic field $b=10^4,$
$b=10^5$, $b=10^6$ and $b=10^{10}$ are, respectively, $C=2.21,\;
9.08, \;31.37$, $\;32.70\times 10^2$, $\;C_\perp = 75.9,\;
2.58\times 10^3,\; 8.38\times 10^4, \;8.49\times 10^{10}$,
$\;C_\parallel =174.3,\; 5.55\times 10^3, \;1.76\times 10^5$,
\;$1.67\times 10^{11}$.

\begin{figure}[htb]
  \begin{center}
   \includegraphics[bb = 0 0 405 210,
    scale=1]{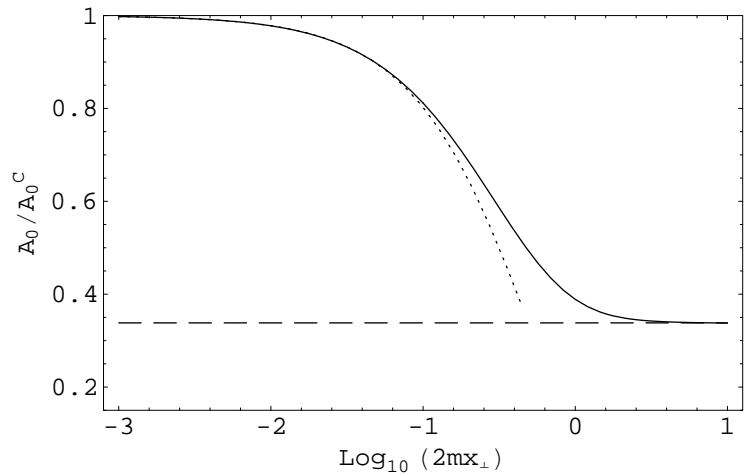}
\caption{Ratio of the modified Coulomb potential (\ref{bessel}) to
the standard Coulomb potential (\ref{free2})
$A_0(0,x_\perp)/A_0^{\rm C}(0,x_\perp)$ drawn as a solid line in
function of the transverse coordinate $x_\perp$ for the magnetic
field value $B=10^4B_0$. The dashed line is the constant $(x_\perp
/x_\perp^\prime)=(1+\alpha B/3\pi B_0)^{-1/2}=0.338$ from the law
(\ref{llargex}). The dotted line is the Yukawa interpolation. 
The abscissa is the same as in Fig.\ref{fig:2}}
\label{fig:3}
  \end{center}
\end{figure}

In Figs. \ref{fig:1} and \ref{fig:2}  functions $A_0(x_3,0)$ and
$A_0(0,x_\perp)$ are drawn by a computer using (\ref{bessel}) for
three values of the magnetic field: $b=10^4,$ $b=10^5$ and
$b=10^6.$ In agreement with the above analysis the curves in Figs.
\ref{fig:1}, \ref{fig:2} approach the Coulomb law $q/4\pi |x_3|$
or $q/4\pi x_\perp$ as $x_3\rightarrow 0$ or $x_\perp\rightarrow
0$, respectively. In Fig. \ref{fig:3} the ratio of the modified
Coulomb law (\ref{bessel}) to the standard Coulomb law
(\ref{free2}) at $x_3=0$ is plotted against the transversal
coordinate $x_\perp$ for the magnetic field $B=10^4B_0$. It
deviates from unity as we move away from the origin $x_\perp=0$.
The dotted line is the interpolation corresponding to the
potential taken in the Yukawa form $ (q/4\pi x_\perp ) \exp
(-2mCx_\perp)$ with $C$ defined by Eq. (\ref{constant}) to be in
this case $C=$2.21. This interpolation is in agreement with the
first two terms of the asymptotic expansion (\ref{expand}). The
Yukawa interpolation is good within the range $0<x_\perp<0.1
(2m)^{-1}=5L_{\rm B}$.

\subsection{Scaling regime}

It is remarkable to note that in the infinite-magnetic-field
limit,  potential (\ref{difference1}), measured in the inverse
Larmour length $L_{\rm B}^{-1}=\sqrt{eB}$ units becomes a
universal, magnetic-field-independent function of coordinates
measured in Larmour units $L_{\rm B}$. To establish this scaling
regime let us make the change of variables in the integral
(\ref{difference2}) $\widetilde{k}_3=k_3L_{\rm B},$
$\widetilde{k}_\perp=k_\perp L_{\rm B}$ and define the new
dimensionless coordinates $x_3=\widetilde{x}_3L_{\rm B},$
$x_\perp=\widetilde{x}_\perp L_{\rm B}.$ After that, we may
substitute unity in place of the function
$T(\widetilde{k}_3^2/4m^2L_{\rm B}^2)$, since it is the value of
this bounded function in the limit $(eB/4m^2)=1/4m^2L_{\rm
B}^2=\infty$. Then Eq.(\ref{difference2})
becomes\bee\label{difference3}\hspace{-3.5cm} \Delta A_0({\bf
x})\simeq \frac {q}{2(2\pi)^2L_{\rm B} }\int_0^\infty
J_0(\widetilde{k}_\perp\widetilde{
x}_\perp)\int_{-\infty}^\infty\left(\frac{\rme^{-\rmi
\widetilde{k}_3\widetilde{x}_3}}{
\widetilde{k}_\perp^2+\widetilde{k}_3^2}-\frac{\rme^{-\rmi
\widetilde{k}_3\widetilde{x}_3}}{
\widetilde{k}_\perp^2+\widetilde{k}_3^2+\frac{2\alpha}{\pi}\exp
\left(-\frac{\widetilde{k}^2_\perp}2\right)} \right)\rmd
\widetilde{k}_3\rmd \widetilde{k}_\perp^2.\eend The
$\widetilde{k}_3$-integration here can be performed by calculating
the residues in the poles on the imaginary axis. Finally one gets
\bee\label{universal}\hspace{-1.5cm}\Delta A_0({\bf x})\simeq 
\frac {q}{4\pi L_{\rm B}}\int_0^\infty
J_0(\widetilde{k}_\perp\widetilde{ x}_\perp)\left(\rme^{-
\widetilde{k}_\perp |\widetilde{x}_3|}
-\frac{\widetilde{k}_\perp\rme^{-|\widetilde{x}_3|\sqrt{\widetilde{k}_\perp^2
+\frac{2\alpha}\pi\exp\left(-\frac{\widetilde{k}_\perp^2}2\right)}}}
{\sqrt{\widetilde{k}_\perp^2+\frac{2\alpha}{\pi}\exp
\left(-\frac{\widetilde{k}^2_\perp}2\right)}}\right)\rmd
\widetilde{k}_\perp.\eend This simple representation can be
further simplified if $x_3$ or $x_\perp$ are large in the Larmour
scale: $|\widetilde{x}_3|\gg 1,$ or $|\widetilde{x}_\perp|\gg 1.$
In this case the integration in (\ref{universal}) is restricted to
the domain $\widetilde{k}_\perp^2\ll 1$ where the exponential
$\exp (-\widetilde{k}_\perp^2/2)$ should be taken as unity. Then
(\ref{universal}) is reduced to \bee\label{yukdif}\Delta A_0({\bf
x})\simeq \frac {q}{4\pi L_{\rm B}}\frac
{1-\exp\{-\left(\frac{2\alpha}{\pi}\right)^{\frac 1{2}}
{\sqrt{\widetilde{x}_\perp^2+\widetilde{x}_3^2}}
\}}{\sqrt{\widetilde{x}_\perp^2+\widetilde{x}_3^2}},\eend which
implies the Yukawa law for (\ref{difference1})
\bee\label{yuk}\hspace{-2cm}A_0({\bf x})\simeq A_0^{\rm Y}({\bf
x})=\frac {q}{4\pi L_{\rm B}}\frac
{\exp\{-\left(\frac{2\alpha}\pi\right)^{\frac 1{2}}
{\sqrt{\widetilde{x}_\perp^2+\widetilde{x}_3^2}}
\}}{\sqrt{\widetilde{x}_\perp^2+\widetilde{x}_3^2}}=\frac
q{4\pi}\frac{\exp\{-\left(\frac{2\alpha b}\pi\right)^{\frac
1{2}}m|{\bf x}|\}} {|{\bf x}|}.\eend
The latter fact can be established if we return  to the previous
representation (\ref{difference3}), which can then be traced back
to (\ref{difference1}), (\ref{difference2}) and finally to
(\ref{bessel}) with
\bee\label{mass}-\kappa_2(0,\infty,0)=\frac{2\alpha}{\pi L_{\rm
B}^2}=m^2\frac{2\alpha b}\pi\equiv M^2\eend substituted for
$-\kappa_2(0,k_3^2, k_\perp^2)$ in the denominator. Here $M$ is
the "effective photon mass" $\;$of Ref.\cite{kuznets}. The
deviation of (\ref{universal}) from (\ref{yukdif}) when
$\widetilde{x}_\perp$ and $\widetilde{x}_3$ are both small is not
important in the sum (\ref{difference1}) against the background of
the first term $A_0^{\rm C}({\bf x})$, singular in the origin
$\widetilde{x}_\perp=\widetilde{x}_3=0$. Therefore, the Yukawa law
(\ref{yuk}) is fulfilled in the vicinity of the charge. By
comparing (\ref{yukdif}) with the second term of the expansion
(\ref{expand}) we deduce that for strong fields\bee\label{C0}
C\sim \frac M{2m}=\sqrt{\frac{\alpha b}{2\pi}}\eend
asymptotically. For the four values of of the external field
$b=10^4,\;10^5,\;10^6,\;10^{10}$ the values of $C$ calculated
following (\ref{C0}) are: 3.41, 10.78, 34.01 and $34.08\times
10^2$. We face the coincidence with its values listed in previous
Subsection, the better the higher the field.

 The Yukawa
law (\ref{yuk}) establishes the short-range character of the
static electromagnetic forces in the Larmour scale. To avoid a
misunderstanding, stress that the genuine photon mass understood
as its rest energy is always exactly equal to zero as a
consequence of the gauge invariance reflected in the
approximation-independent relation $\kappa_a(0,0,0)=0.$
Correspondingly, the potential, produced by a static charge,
should be long-range for sufficiently large distances. This is the
case, indeed. The point is that the scaling regime
(\ref{universal}), as it follows from its derivation, is valid in
the limit $eB/4m^2=\infty$. In this limit the ratio of the Compton
length $(2m)^{-1}$ to the Larmour length $L_{\rm B}=(eB)^{-1/2}$
is infinite. Therefore, Eqs. (\ref{universal}), (\ref{yuk}) cannot
be extended to the distances of Compton scale and larger, where
the long-range character of the electromagnetic interaction is
restored, as we shall see in the next Section.

\section{Large-distance behavior}

\subsection{Large $x_\perp$ in Larmour scale}

For large transverse distances the term linearly growing with the
magnetic field (\ref{2}) 
leads to suppression of the static potential in the transverse
direction.

To be more precise, consider the region
\bee\label{largexperp}x_\perp\gg \frac
{m^{-1}}{\sqrt{2b}}=\frac{L_{\rm B}}{\sqrt{2}}.\eend Once the
Bessel function $J_0$ in (\ref{bessel}) oscillates and decreases
for large values of its argument $k_\perp x_\perp$, the main
contribution into the integral over $k_\perp^2$ in (\ref{bessel})
comes from the integration variable domain $k_\perp^2\ll 2m^2b$,
and the dependence upon $k_\perp^2$ in $\kappa_2$ may thus be
disregarded. Then the $k_\perp^2$-integration in (\ref{bessel})
can be explicitly performed to give (we use Eq.6.532.4 of the
reference book \cite{ryzhik})
\bee\label{largexperp2}A_0(x_3,x_\perp)\simeq \frac {2q}{(2\pi
)^2}\int_0^{\infty}
\mathcal{K}_0\left(x_\perp\sqrt{k_3^2-\kappa_2(0,k_3^2,0)}\right)\cos
(k_3x_3)\rmd k_3,\eend where $\mathcal{K}_0$ is the McDonald
function of order zero, and
\bee\label{kappa00}\kappa_2(0,k_3^2,0)=-\frac{2\alpha b}\pi
m^2T\left(\frac {k_3^2}{4m^2}\right).\eend 
As the McDonald function $\mathcal{K}_0$ decreases exponentially
when its argument increases, only small values of the square root
contribute into integral (\ref{largexperp2}), 
and the $k_3$-integration domain in it is restricted to the
interval $k_3^2\ll 4m^2$, wherein \bee\label{T(0)}
T\left(\frac{k_3^2}{4m^2}\right)\simeq \frac{k_3^2}{6m^2}.\eend
Then the potential form (\ref{largexperp2}) becomes (we use Eq.
6.671.14 of the reference book \cite{ryzhik})
\bee\label{llargex}A_0(x_3,x_\perp)\simeq \frac
{2q}{(2\pi)^2}\int_0^\infty\mathcal{K}_0\left(x_\perp
k_3\sqrt{1+\frac{\alpha b}{3\pi}}\right)\cos (k_3x_3)\rmd
k_3=\nonumber\\=\frac 1{4\pi}\frac
{q}{\sqrt{(x^\prime_\perp)^2+x_3^2}},\quad\qquad
x^\prime_\perp=x_\perp\left(1+\frac{\alpha
b}{3\pi}\right)^{1/2},~~x^\prime_\perp>x_\perp.\eend 

Eq.(\ref{llargex}) is an anisotropic Coulomb law, according to
which the attraction force decreases with distance from the source
along the transverse direction faster than along the magnetic
field (remind that $b\equiv (B/B_0)\gg 1$), but remains
long-range. The equipotential surface is an ellipsoid stretched
along the magnetic field. The electric field of the charge  ${\bf
E}=-{\bf \nabla}A_0(x_3,x_\perp)$ is a vector with the components
$(q/2\pi)(x_3^2+\beta^2x_\perp^2)^{-3/2}(x_3,\beta^2{\bf
x}_\perp)$, where $\beta=(1+\alpha b/3\pi)^{1/2}$. It is not
directed towards the charge, but makes an angle $\phi$ with the
radius-vector $\bf r$,
cos$\phi=(x_3^2+\beta^2x_\perp^2)(x_3^2+\beta^4x_\perp^2)^{-1/2}
(x_3^2+x_\perp^2)^{-1/2}.$ In the limit of infinite magnetic
field, $\beta=\infty$, the electric field of the point charge is
directed normally to the axis $x_3$.  This regime corresponds to
the dielectric permeability of the vacuum independent of the
frequency, with its dependence on $\bf k$ (spatial dispersion)
being reduced solely to that upon the angle in the space ($cf,$
\cite{zhetf}). 

 The result (\ref{llargex}) is in
agreement with  the  curves in Figs.\ref{fig:2},
\ref{fig:3} drawn for $x_3=0$. 
The curves in Fig.\ref{fig:2}, approach the standard Coulomb
law when $x_\perp\rightarrow 0$ 
in accordance with (\ref{expand}), as explained above,  rather
sharply fall down, following the Yukawa law (\ref{yuk}) in the
Larmour range $x_\perp\sim L_{\rm B}$ to  reach the asymptotic
long-range regime $A_0(0,x_\perp)\simeq q/4\pi x^\prime_\perp$ for
larger $x_\perp$ in the region (\ref{largexperp}).
\subsection{Large $x_3$}

It remains to consider the remote coordinate region of large
$x_3$, complementary to ({\ref{largexperp}).

To this end we apply the residue method to the inner integral over
$k_3$ in (\ref{bessel}).
 Using the
integral representation (\ref{T})  the function $\kappa_2$
(\ref{2}) may be, for a fixed positive value of $k_\perp^2$,
analytically continued from the real values of the variable  $k_3$
into the whole complex plane of it, cut along two fragments of the
imaginary axis. In the lower half-plane the cut runs from Im$~
k_3=-2m$ down to Im $k_3= -\infty$, while in the upper half-plane
it extends within the limits $2m \leq {\rm Im}~ k_3\leq\infty$.
Other singularities of the $k_3$-integrand in (\ref{bessel}) are
poles yielded by zeros of the denominator, i.e. solutions of the
equation (associated with the photon dispersion
equation)\bee\label{dispersion}
k_\perp^2+k_3^2-\kappa_2(0,k_3^2,k_\perp^2)=0.\eend  As $k_\perp$
varies within the limits $(0,\infty)$ two roots of this equation
$k^\pm_3=\pm \rmi K(k_\perp)$ move along the imaginary axis from
the point $K(0)=0$ to the points  $k^\pm_3=\pm \rmi K(\infty)=\pm
\rmi 2m$ \cite{nuovcimlet}\cite{annphys},\cite{zhetf}. There is
yet another branch of the solution to equation (\ref{dispersion}),
corresponding to the photon absorption via the $\gamma\rightarrow
e^+e^-$-decay, but the corresponding poles lie in the nonphysical
sheet of the described complex plane, behind the cuts, and will
not be of importance for the consideration below.

Let us consider positive values of $x_3$. Negative values can be
handled in an analogous way. Turning the positive part of
integration path $0\leq k_3\leq\infty$ clockwise to the lower
half-plane by the angle $\pi/2$, and the negative part
$-\infty\leq k_3\leq 0$ counterclockwise by the same angle, and
referring to the fact that the exponential $\exp (-\rmi k_3x_3)$
decreases, for $x_3>0$, in the lower half-plane of $k_3$ as
$|k_3|\rightarrow\infty$ so that the integrals over the remote
arcs may be omitted, we get a representation for the inner
integral in
(\ref{bessel})\bee\label{inner}\int_{-\infty}^\infty\frac{\rme^{-\rmi
k_3x_3}\rmd k_3}{
k_\perp^2+k_3^2-\kappa_2(0,k_3^2,k_\perp^2)}=\nonumber\\=\rmi
\int_{2m}^\infty\rme^{-|k_3|x_3} \Delta(|k_3|^2,k_\perp^2)\rmd
|k_3| -2\rmi \pi \rme^{-(K(k_\perp^2)x_3)}{\rm
Res}(k_\perp^2),\eend where Res$(k_\perp^2)$ designates the
residue of the expression $D_2 (0,-|k_3|^2,k_\perp^2)=\left(
k_\perp^2-|k_3|^2-\kappa_2(0,-|k_3^2|,k_\perp^2)\right)^{-1}$ in
the pole $k_3^-=-iK(k^2_\perp)$, while
$\Delta(|k_3|^2,k_\perp^2)=D_2(0,-|k_3|^2+\rmi 0,k_\perp^2)-
D_2(0,-|k_3|^2-\rmi 0,k_\perp^2)$ is the cut discontinuity. It was
explained above that $0<K(k_\perp^2)< 2m$ everywhere but in the
limit $k_\perp\rightarrow\infty$, where $K=2m$. Consequently the
residue term in (\ref{inner}) dominates over the cut-discontinuity
term everywhere in the $k_\perp$-integration domain in the outer
integral in (\ref{bessel}), except for the region near $k_\perp=
\infty$. In this limit, however, $\kappa_2$ disappears due to the
exponential in (\ref{2}), together with the cut discontinuity,
since the latter is only due to the branching points in the
function (\ref{T}). Therefore, keeping the residue term in
(\ref{inner}) as the leading one, we neglect the contribution that
decreases with large longitudinal distance at least as fast as
$\exp (-2m|x_3|).$ In this way we come to the asymptotic
representation of the potential (\ref{bessel}) in the region of
large longitudinal distances $|x|_3\gg (2m)^{-1}$ (the negative
$x_3$ at this step are also included - to treat them the fragments
of integration path were rotated in the directions opposite to the
above) \bee\label{residue}A_0({\bf x})\simeq \frac
{q}{8\pi}\int_0^\infty \frac{J_0(k_\perp
x_\perp)}{K(k^2_\perp)(1+H(-K^2(k^2_\perp),k_\perp^2)}
\rme^{-(K(k^2_\perp)x_3)}\rmd k^2_\perp,\eend where \bee\label{H}
H(k_3^2,k_\perp^2)=\frac{2\alpha bm^2}\pi\exp
\left(-\frac{k_\perp^2}{2m^2b}\right)\frac\rmd{\rmd k_3^2}
T\left(\frac{k_3^2}{4m^2}\right). \eend Here $K^2(k_\perp^2)$ is
the solution of equation (\ref{dispersion}) in the negative region
of the variable $k_3^2$ - see \cite{zhetf} for its form.
$K(\infty)= 2m$, $K(0)=0$. $T$ is given by (\ref{T}).

Due to the exponential factor in the integrand of (\ref{residue}),
for large $x_3$ the main contribution  comes from the integration
region of $k_\perp$ that provides minimum to the function
$K(k_\perp)$. The minimum value of $K(k_\perp)$ is  zero. It is
achieved in the point $k_\perp=0$ - a manifestation of the fact
that the photon mass defined as its rest energy is strictly equal
to zero owing to the gauge invariance: $\kappa_a(k_0=k_3={\bf
k}=0)=0$. In view of (\ref{kappa00}) and (\ref{T(0)}), near the
point $k_\perp=0$ the dispersion equation (\ref{dispersion}) has
the solution $K(k_\perp)=k_\perp/\sqrt{1+\alpha b/3\pi}$.
Simultaneously, near the minimum point $1+H(0,0)=1+\alpha b/3\pi$.
With these substitutions and the use of 6.661.1 of \cite{ryzhik},
Eq. (\ref{residue}) becomes again the anisotropic Coulomb law
(\ref{llargex}) $(q/4\pi)/((x^\prime_\perp)^2+x_3^2)^{1/2}$. We
have, therefore, established its validity everywhere in the region
remote from the center, irrespectively of the direction.
\begin{figure}[htb]
  \begin{center}
   \includegraphics[bb = 0 0 405 210,
    scale=1]{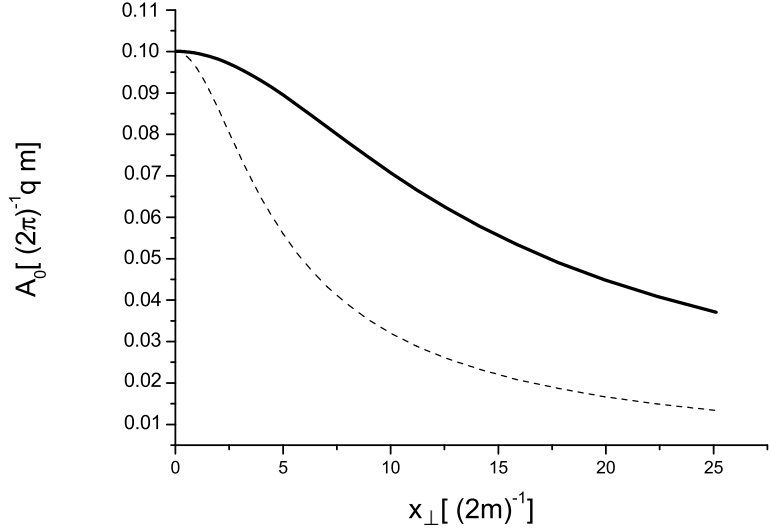}
\caption{Modified Coulomb potential (\ref{residue}) plotted
against the transverse coordinate $x_\perp$ with  the longitudinal
coordinate fixed at the large value $x_3=10(2m)^{-1}$. The dashed
line corresponds to the magnetic field  value $B=10^4B_0$. Solid
line is the standard Coulomb law (\ref{free2}) $A_0^{\rm
C}(x_3,x_\perp)=qm/2\pi ((2mx_\perp)^2+100)^{-1/2}$. The dashed
line is indistinguishable   from the anisotropic Coulomb law
(\ref{llargex}) in the scale of the drawing.  The coordinate axes
are the same as in Fig.\ref{fig:2}} \label{fig:4}
  \end{center}
\end{figure}
\begin{figure}[htb]
  \begin{center}
   \includegraphics[bb = 0 0 405 210,
    scale=1]{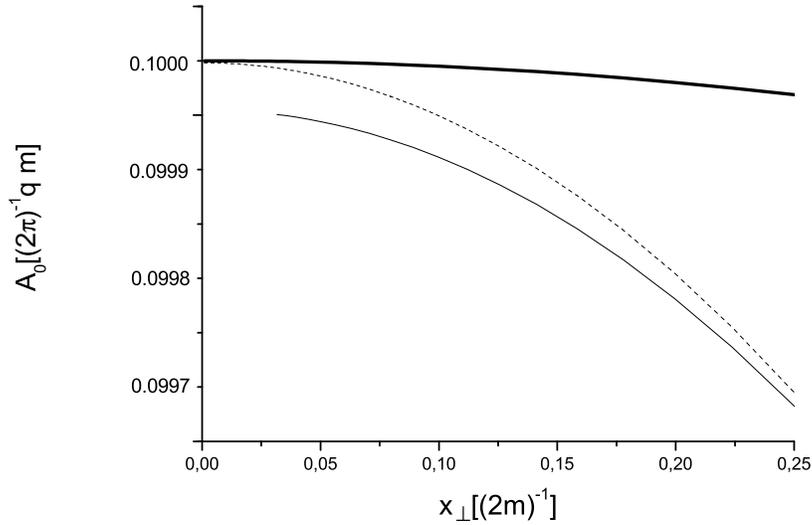}
\caption{The same as Fig.\ref{fig:4}, but viewed at a detailed
scale near $x_\perp=0$. The thin solid line is the anisotropic
Coulomb law (\ref{llargex}) $qm/2\pi ((0.338\cdot
2mx_\perp)^2+100)^{-1/2}$} \label{fig:5}
  \end{center}
\end{figure}
In agreement with this result the curves in Figure 1 for
$A_0(x_3,0)$ are approaching the Coulomb law $q/4\pi |x_3|$ as
$x_3$ grows. To see how fast they are doing this, set $x_\perp=0$
in the difference (\ref{difference2}). By replacing the
exponential in the expression for $\kappa_2$ (\ref{2}) by unity we
overestimate $|\Delta A_0(x_3,0)|$. Then the  integration over
$k_\perp^2$ can be explicitly performed to yield the inequality
\bee\label{speed1}|\Delta A_0(x_3,0)|< \left|\frac
q{8\pi^2}\int_{-\infty}^\infty \rme ^{-\rmi
x_3k_3}\ln\left(1+\frac {2\alpha bm^2}{\pi
k_3^2}T\left(\frac{k_3^2}{4m^2}\right)\right)\rmd k_3\right|.\eend
The only singularities of the integrand of (\ref{speed1}) as a
function of the complex variable $k_3$ are (inverse square root)
branchings on the imaginary axis in the points  $\pm\rmi 2m$ owing
to the analytical properties of the function $T(y)$ (\ref{T}).
Note that the argument of the logarithm in (\ref{speed1}) may
disappear only on the second sheet of the complex plane. By
turning the integration path to the lower half-plane, when $x_3$
is positive, or to the upper half-plane otherwise, we reduce
(\ref{speed1}) to an integral of the cut discontinuity multiplied
by $\exp (x_3{\rm Im}k_3)$ along the imaginary axis of $k_3$, the
minimum integration value of $|{\rm Im}k_3|$ being $2m$. Therefore
the difference (\ref{difference2}) between the potential
$A_0(x_3,0)$ and its large-$x_3$ asymptote $q/4\pi|x_3|$ decreases
in Figure 1 at least as fast as exp$(-2m|x_3|)$.

Eq.(\ref{residue}) was used for computer calculation with the
large value $x_3=10m^{-1}$. It has  lead to the curve shown in
Figs \ref{fig:4}, \ref{fig:5}. In the region  (\ref{largexperp})
it agrees
with the result (\ref{llargex}), valid  in that region 
($L_{\rm B}=0.02 (2m)^{-1}$ for $b=10^4$). In practice
(\ref{residue}) and (\ref{llargex}) are the same.

\section {Conclusion}
The modification of the Coulomb law should affect, first of all,
the field of an atomic nucleus, placed in a magnetic field. 
Properties of matter (including individual atoms)
 at the surface of strongly magnetized neutron stars and various
physical processes (such as radiation of particles) where the
electric field of particles is important (for a review on physics
of strongly magnetized neutron stars, see \cite{HD06}) may become
sensitive to the present modification of the Coulomb law.
 \ack This work was
supported by the Russian Foundation for Basic Research (project no
05-02-17217) and the President of Russia Programme
(LSS-4401.2006.2), as well as by the Israel Science Foundation of
the Israel Academy of Sciences and Humanities.

\end{document}